# Trapped low magnetic field dynamics in YBCO single crystals

V.Yu.Monarkha, A.A.Shablo, and V.P.Timofeev

*B.Verkin Institute for Low Temperature Physics & Engineering National Academy of Sciences of Ukraine, 47, Lenin Ave., 61103 Kharkov, Ukraine*
*E-mail: timofeev@ilt.kharkov.ua*

**Abstract.** Trapped low magnetic flux dynamics including local ones are investigated in YBCO single crystals in the strong thermal fluctuations domain near the superconducting phase transition temperatures. The essential difference from quasi-logarithmic isothermal relaxation of a magnetization behavior (observed earlier in high magnetic fields) was established. Within the framework of classic model of a thermal activated vortices creep the estimation of an effective pinning potential was made at these temperatures.

The main part of theoretical and experimental works on the magnetic flux dynamics and structure studies in high temperature superconductors (HTSC) are concentrated on revealing mechanisms that determine maximal current carrying ability of the superconductor (SC) in strong magnetic fields [1,2].
Research of the magnetic fluxes behavior in weak magnetic fields (close to earth magnetic field density) and at temperatures close to critical ones, are especially topical when developing high-sensitive HTSC SQUIDs and derivative liquid nitrogen-cooled electronics. The magnetic flux dynamics is associated with creep and vortices hopping. It depends on vortices pinning on structural defects and is determined by the thermal activation energy of this process. The vortex hopping makes a major contribution to superconductive elements self-noise, and depends on the presence and effectiveness of the pinning centers in the HTSC materials.

It is known, that in HTSC SQUIDs, there exists two dominating noise sources: the fluctuation of the critical current in the Josephson's junction, and vortices motion in the HTSC material of the sensor [3]. The probability of the vortex hopping increases exponentially with the temperature and effective pinning energy. That's why the HTSC microstructure and appropriate activation energies are the main factors that determine the sensitivity of the HTSC SQUIDs. The main goal of this work is to study the magnetic flux dynamics in $YBa_2Cu_3O_{7-x}$ (YBCO) single crystals with different structure in low constant magnetic fields, at temperature range close to the phase transition region.

While this research the isothermal relaxation of the magnetic momentum of sample produced by Meissner currents or trapped magnetic fluxes was measured by means of a SQUID-based magnetometery. This method provides the sensitivity needed and allows reducing the preparation procedure with the sample. The used SQUID sensitivity allows considerably to decrease the applied fields (dawn to $H$ = 0.01-0.1 Oe), and even to detect spontaneous currents and magnetic moments.

As the main objects for study we selected pure *c*-oriented YBCO single crystal samples [4]. After the annealing in oxygen at 400 °C required for optimization of doping, the single crystals had the typical critical temperature $T_c$ =93 K, measured by the resistive method in a "zero" magnetic field. The superconducting transition width is about 0.3 K that points to the high quality of the samples. The annealing leads to transformation of the tetragonal crystal structure into the orthorhombic one and, as a result, to the formation of the twin boundary (TB) planes. To investigate the influence of these plane defects on the pinning processes, we have selected two types of YBCO single crystals with sizes close to 1×1×0.02 mm. The first group had unidirectional twin boundaries oriented parallel to the *c*-axis of the crystal through the whole crystal, and the others contained blocks of twin boundaries with different orientation.

The resistive state in HTSC originates from the beginning of magnetic vortices motion, when the acting Lorentz forces start to exceed pinning ones. Under the influence of these forces, and also under effect of thermal activation hopping, happened with probability $\sim \exp(-U/kT)$, the vortices start to move, the dissipation of energy appears (here: $U$ - effective activation energy of vortices hopping, equals to average depth of pinning potential; $k$ - Boltzmann constant; $T$ - temperature). These processes also determine value of a superconductive critical current ($I_c$).

The ideal superconductor, placed in a weak magnetic field, should be in Meissner condition. In the realistic SC of the final sizes at the presence of surface and volume defects and at temperatures, close to critical one, the magnetic field starts to penetrate inside a sample even at $H << H_{c1}$ ($H_{c1}$ – the first critical field of ideal flawless superconductor of the ellipsoidal shape). The thermal activating creep of vortices results in reallocating and damping of supercurrents, relaxation of magnetization ($M$) in time.

The data on research of a magnetic relaxation in superconductors can be used for obtaining main parameters of the vortexes pinning mechanism. Thus in the simplest case the effective depth of a pinning potential can be estimated from measurement of a normalized isothermal relaxation rate

$S = 1/M_0 (dM/d \ln t) = - kT/U,$ (1)

where $M_0$ - initial value of a magnetization, for which one, as a rule, take a magnetization in a Bean's critical state regime [2]. However practically all published results on a magnetic relaxation of HTSCs were performed in strong magnetic fields (hundred Oe or even several kOe), when the essential role are played complex processes in rigid, well formed lattice of magnetic vortices. The experimental data received are very sensitive to the magnetic field orientation in relation to the main crystallographic planes of the sample.

As it was shown in [5], according to the theory of collective pinning in weak magnetic fields, creep of noninteracting vortices will be realized, the average velocity of the flux lines does not depend on a magnetic field value, and observed data are less sensitive to the magnetic field deviation from the direction of *c*-axis of HTSC sample. The research of samples magnetization and magnetic relaxation in the superconducting phase transition region in weak magnetic fields are made with the help of a SQUID-based magnetic susceptibility meter. The standard technique for $M(T,t)$ measurement in a homogeneous DC magnetic field of the solenoid was utilized. The residual magnetic field of the Earth in the working area was shielded and did not exceed 0.5 mOe.

It is allowed one to produce a sample cooling and transferring it in superconducting state on a zero field cooling (ZFC) method that is preferential at research of $M(T)$. At the analysis of $M(t)$ a sample cooling down to temperature of $\approx$ 77 K was made in the chosen magnetic field, that is on a field cooling (FC) method, then the $T$ set was selected and fixed, the field was turn off, and the behavior of $M(t)$ was recorded.

Fig.1 shows the normalized rate the relaxation of the magnetization $S$ due to trapped fluxes for one of studied YBCO single crystals ($T_c$=93.5 K) with unidirectional TBs at temperatures closed to the SC transition region. The magnetic field of the solenoid was parallel to *c* axis of single crystal and equals to 120 A/m ($\approx$ 1.5 Oe) in value. At such orientation magnetic field is parallel to the TB planes, and Abrikosov vortices are pinned most effectively.

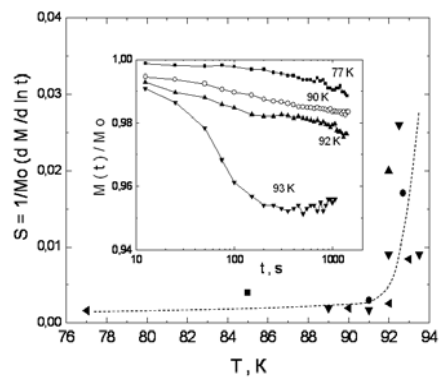

Fig.1. Magnetization normalized rate of YBCO single crystal in the region near $T_c$. Inset: A nonlogarithmic $M(t)$ behavior.

Insert in Fig. 1 shows the magnetization dynamics of this single crystal in time $M(t)$. As you can see at the beginning of the relaxation process the magnetization doesn't change in time, or changes very slowly. At relatively small temperatures it can be explained by the exponential diminution of the magnetic flux thermal creep, and by the presence of the random Josephson's links in the twin boundaries area. For temperatures closed to $T_c$ magnetization has an initial nonlogarithmic shape. Then it behavior can be described by Anderson-Kim flux-creep model, and at long times magnetization demonstrates some saturation with the presence of strong thermal fluctuations.

The trapped flux dynamics is also connected with the Bin-Livingston potential barrier. To avoid the influence of the uncontrolled edge effects on the flux relaxation, a series of experiments using a tiny solenoid has been made. The diameter of this solenoid was 0.2mm, which is significantly smaller then the sample dimensions in $ab$-plane ($\approx 1.4 \times 1.2$ mm$^2$). This allows decreasing the edge barrier influence on the magnetic flux dynamics.

Using the normalized magnetic relaxation rate data, an effective $U$ depth has been estimated (See Fig.2).

The results of the experiments with both sample type give a reason to believe that the twinning planes provide conditions for the Josephson's nets with randomly distributed parameters to appear. The twin boundaries include $CuO_x$ layers that contain oxygen vacancies. The oxygen vacancies cause the suppression of the superconductive order parameter, what leads to reduction the vortex energy captured by the TBs. That's why the density of the vortices on the twin boundaries can be greater than anywhere in the crystal.

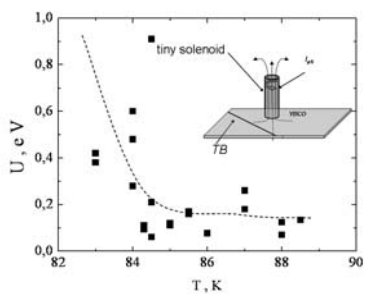

Fig.2. Effective pinning potential of tested YBCO sample. Inset: main geometry of experiment.

Thus, for the first time isothermal magnetic relaxation of HTSC samples with different crystalline pattern was investigated in very weak DC fields ($H \leq 1$ Oe) at temperatures, close to critical ones. It was also studied the trapped local field dynamics in temperature range from $T \approx 77$ K, and to the temperatures that are close to $T_c$. Significant influence of the crystal structure of the samples on the magnetization relaxation rate has been shown, and the possibility of the significantly nonlogarithmic relaxation behavior in presence of the strong thermal fluctuations has been determined. In the framework of the thermal activated creep model effective pinning potential estimation has been made for single crystals with unidirectional twin boundaries.

The results obtained are useful to understand HTSC pinning mechanisms, can be used while constructing high sensitive superconducting devices, to decrease magnetic self-noise and to increase sensitivity of the nitrogen-cooled superconductive sensors.

The authors are grateful to M.A. Obolenskiy and to A.V. Bondarenko for the single crystal samples given.